\documentclass[11pt,a4paper]{article}

\usepackage[margin=2.8cm,bottom=3.5cm]{geometry}

\usepackage[margin=2.8cm,bottom=3.5cm]{geometry}

\usepackage{amsmath, amssymb}
\usepackage{color}
\pdfoutput=1 

\usepackage{graphicx} 
\usepackage{float}
\usepackage{subcaption}
\usepackage[T1]{fontenc} 

\usepackage{color}
\pdfoutput=1 

\usepackage{graphicx} 
\usepackage{subcaption}
\usepackage[T1]{fontenc} 

\usepackage[backend=biber, sorting=none, style=numeric-comp]{biblatex}
\title{Scattering cross-section in gravitating cosmic string spacetimes}

\author{
        Marcos Silva \\
        Departamento de F\'isica, Universidade Federal de Pernambuco,\\
        Av. Prof. Moraes Rego, 1235, Recife - PE - 50670-901, Brazil\\
        marcos.viniciussantos@ufpe.br
            \and
        Azadeh Mohammadi\\
        Departamento de F\'isica, Universidade Federal de Pernambuco,\\
        Av. Prof. Moraes Rego, 1235, Recife - PE - 50670-901, Brazil\\
        azadeh.mohammadi@ufpe.br
}

\begin{document} 
\maketitle

\begin{abstract}
We propose a modification of the partial wave approach to deal with the relativistic quantum scattering of bosonic and fermionic particles in a class of models concerning gravitating cosmic string spacetimes. These spacetimes are characterized by the Minkowski line element at the center of the vortex, non-vanishing curvature at a finite distance from the center but with a conical structure far from the core. We find the correction in the partial wave expansion and the phase shift. Consequently, we show the explicit form of the scattering amplitude and the correction to the differential cross-section for a massive scalar field. We also implement our formalism in a toy model mimicking this class of gravitating cosmic string spacetime. Moreover, we discuss the procedure to apply this formalism to a massive Dirac field.
\end{abstract}

\section{Introduction}
In 1976, Kibble showed that cosmological phase transitions leading to distinct non-zero vacuum states of the Higgs field should, in principle, cause the formation of topological defects, namely domain walls, monopoles, and cosmic strings \cite{kibble1976topology}. Among these defects, cosmic strings are the most promising ones. Ever since, many cosmic strings solutions have been found in models with spontaneous symmetry breaking, including Nielsen-Olesen vortex in the Abelian-Higgs model in 1973 \cite{nielsen1973vortex}, Semilocal strings by Vachaspati and Achucarro in 1991 \cite{vachaspati1991semilocal}, and electroweak strings in the Weinberg-Salam model by Vachaspati in 1993 \cite{vachaspati1993electroweak}. For a thorough review of the field theoretical and cosmological aspects of various types of cosmic string solutions, see \cite{achucarro2000semilocal,sakellariadou2007cosmic}.

The gravitational aspects of cosmic strings with both wire and non-wire approximation have been extensively studied considering different matter models \cite{slagter1998self, garfinkle1985general, laguna1989spacetime, christensen1999complete, dyer1995complete, linet1987vortex, Brihaye_Lubo, vilenkin1981gravitational, Antonio_Eugenio}. In these models, in order to find the metric for a gravitating cosmic string, one needs to couple a Higgs potential with a stable topological solution to gravity.
In most cases, the metric is found numerically. Although there are various possibilities for the Higgs model leading to such solutions, one of their common features is a conical spacetime far from the vortex core.\
Conical structures \emph{per se} may create many exciting features in the dynamics of fields. For example, they can affect the vacuum expectation values of the physical observables. An ideal cosmic string can induce vacuum polarization as well as bosonic and fermionic charge and current densities (see for example  \cite{bragancca2020induced, mohammadi2016finite} and references therein). Not surprisingly, the scattering problem shall present rich features as well. In \cite{jackiw1985lower} it was shown how the scattering of two massive point particles is related to the scattering of one particle on a cone, and in \cite{deser1988classical} the authors analyzed the classical and quantum non-relativistic scattering of a particle in a conical space. One of the main results in \cite{deser1988classical} is that the conical geometry creates singularities in the scattering amplitude and consequently in the scattered wavefunction if one follows the canonical procedure in quantum mechanics. The authors solved this issue by modifying the asymptotic incident wave. Motivated by this approach, \cite{barroso2017quantum} showed that the evolution of a wave packet in the conical spacetime of an ideal cosmic string is highly dependent on the many possible boundary conditions of the scalar field at the vertex of the cone.  
In \cite{spinelly2001relativistic}, the authors generalized the procedure proposed in \cite{deser1988classical} to the relativistic scattering of a charged scalar field by an ideal cosmic string considering non-minimal coupling to gravity, and in \cite{neto2020scalar} they analyzed the scattering of scalar particles in the same spacetime but considering minimal and non-minimal coupling to vector and scalar fields.\
In the present work, we propose a new formalism dealing with the scattering problem of bosonic and fermionic fields in the general spacetime of a gravitating cosmic string, which possesses a conical structure far from the core. In this formalism, in order to find the scattering amplitude and differential cross-section, we have also modified the asymptotic incident wave, although with a different approach from the one in \cite{deser1988classical}.

The structure of the paper is as follows. In section \ref{sec1}, we study the main aspects of a gravitating cosmic string spacetime, while in section \ref{sec2}, we derive the formula for the scattering amplitude of a massive scalar field and discuss novel aspects of our approach. We also give a simple example, in section \ref{sec6}, to show the utility of the new approach. In section \ref{sec4}, we comment on how this approach is employed in the fermionic case. Finally, in section \ref{sec5}, we present our conclusions. Throughout the paper we use natural units ($\hbar = c = 1$).

\section{Conical structure}
\label{sec1}
One of the simplest conical structures is an ideal cosmic string, or equivalently a "confined" vortex represented by the line element
\begin{equation}
    ds^2 = dt^2 - dr^2 - b^2 r^2 d\varphi^2 - dz^2,
\end{equation}
with $b = 1 - 4\mu$, $\mu$ being the string's mass per unit length. It is a locally flat and cylindrically symmetric spacetime.  Defining $b \varphi$ as the angular coordinate, one can see that it does not cover the whole $2\pi$ for $b<1$, which clearly gives origin to the conical structure. 
However, the conical structure also appears when one deals with an "open" or extended vortex, in which the energy density is spread over space. In this case, the most general form of the line element respecting the cylindrical symmetry is as follows
\begin{equation}
    ds^2 = N^2(r)dt^2 - dr^2 - L^2(r)d\varphi^2 - N^2(r)dz^2.
    \label{metric_general}
\end{equation}
It presents a flat spacetime in two regions although, in general, with different parametrizations. When $r \rightarrow 0$, the metric coefficients are 
\begin{equation}
    \begin{gathered}
    N(r) \rightarrow 1\\
    L(r) \rightarrow r,
    \end{gathered}
    \label{metric_zero}
\end{equation}
and when $r \rightarrow \infty$ we have
\begin{equation}
    \begin{gathered}
    N(r) \rightarrow a\\
    L(r) \rightarrow br + c,
    \end{gathered}
    \label{metric_infty}
\end{equation}
where $a$, $b$ and $c$ are constants in spacetime. However, they may depend on the parameters of the model for the vortex. In both limits, $r \rightarrow 0$ and $r \rightarrow \infty$, the Ricci tensor and consequently the curvature are zero, representing a flat spacetime. Though, in between, $0 < r < \infty$, the metric can be arbitrarily complicated and, in general, presents non-vanishing curvature.  Concrete examples of this behavior can be found in \cite{Brihaye_Lubo} for a gravitating cosmic string originating from the Abelian-Higgs model, and in \cite{Antonio_Eugenio} for a non-abelian extension with two bosonic sectors, taking the line element (\ref{metric_general}).

\section{Scalar field scattering}
\label{sec2}
One of the most relevant physically observable quantities, especially in particle physics, is the cross-section of particle scattering. In this section, our goal is to find the phase shift and the corresponding differential cross-section of a scalar field in a gravitating cosmic string spacetime background. Let us start with the Klein-Gordon equation non-minimally coupled with gravity
\begin{equation}
    (\Box + M^2 + \xi R)\Phi = 0,
    \label{scalar_kleingordon}
\end{equation}
where $R$ is the Ricci scalar, $M$ is the mass, and $\xi$ is the non-minimal coupling.
Replacing the d'Alembertian operator by $\Box=\frac{1}{\sqrt{-g}}\partial_\mu{(\sqrt{-g}g^{\mu\nu}\partial_\nu{\Phi})}$, we get
\begin{equation}
    \Bigg{\{}\frac{1}{N^2L}\Bigg{[}L\partial_{t}^2-\Big{(}2NN'L+N^2L'\Big{)}\partial_{r}-N^2L\partial_{r}^2-\frac{N^2}{L}\partial_{\varphi}^2-L\partial_{z}^2\Bigg{]} + M^2 +\xi R\Bigg{\}}\Phi=0,
    \label{scalar_eom1}
\end{equation}
where the prime denotes the derivative with respect to $r$. Taking into account the cylindrical symmetry and energy conservation, we substitute the following ansatz
\begin{equation}
    \Phi = e^{\mp iEt} e^{ikz} \sum_{m=-\infty}^{\infty} a_m R_{m}(r) e^{im\varphi},
    \label{scalar_solution_ansatz}
\end{equation}
into the differential eq. (\ref{scalar_eom1}) and define $\lambda^2 \equiv E^2 - M^2 - k^2$ which result in
\begin{equation}
    R_{m}''(r) + \left(\frac{2N'(r)}{N(r)}+\frac{L'(r)}{L(r)}\right)R_{m}'(r)+\left[\frac{\lambda^2}{N^2(r)}- M^2\left(1-\frac{1}{N^2(r)}\right)-\frac{m^2}{L^2(r)}+ \xi R\right]R_{m}(r)=0.
    \label{scalar_eom2}
\end{equation}
The summation over $m$ and the parameter $a_m$ reflects the fact that for a linear equation, the sum of the solutions with any arbitrary constant prefactor is still a solution.

The solution to eq. (\ref{scalar_eom2}) near the origin is a Bessel function in the form $R_m(r \rightarrow 0) \propto J_m(\lambda r)$, where we absorb the proportionality constant in $a_m$. If we impose the solution near the origin to be a plane wave in the x-direction, the coefficients are automatically determined, $a_m = i^m$.
The solution in the limit $r \rightarrow \infty$, where the metric components converge to eq. (\ref{metric_infty}), is $R_m(r \rightarrow \infty) = b_m J_{m^\prime}(\lambda^\prime w) + c_mY_{m^\prime}(\lambda^\prime w)$. In this solution, $Y_m$ is the Neumann function of order $m$, $w = r + c/b$, $m^\prime = m/b$ and ${\lambda^\prime}^2 = \lambda^2/a^2 - M^2(1 - 1/a^2)$. Taking the assymptotic form of both Bessel and Neumann functions together with $b_m = C_m \cos d_m$ and $c_m = -C_m \sin d_m$, leads to 

\begin{equation}
    R_m(r \rightarrow \infty) = C_m \sqrt{\frac{2}{\pi \lambda^{\prime} r}} \cos \left(\lambda^{\prime} r + \beta_{m^\prime}\right),
    \label{scalar_asymp_solution}
\end{equation}
where $\beta_{m^\prime} = \frac{\lambda^\prime c}{b} - \alpha_{m^\prime} + d_m(\lambda)$, $\alpha_m = \frac{\pi}{2}(m + 1/2)$ while $C_m(\lambda)$ and $d_m(\lambda)$ are model-dependent constants to be determined, usually numerically. It is important to notice that the phase shift of the m-th mode is given by $\delta_m(\lambda) = \beta_{m^\prime} + \alpha_m = \frac{\lambda^\prime c}{b} + \frac{m\pi}{2}\left(1 - \frac{1}{b} \right) + d_m(\lambda)$. 
Also, note that if the flat spacetime for $r \rightarrow 0$ and r $\rightarrow \infty$ is with the same parameterization, i.e. $c=0$ and $b=1$, the phase shift becomes equal to $d_m(\lambda)$ determined by the form of the potential in $0<r<\infty$.  Moreover, the non-minimal coupling $\xi$ becomes irrelevant in the general form of the solutions since the curvature vanishes in both limits, $r \rightarrow 0$ and $r \rightarrow \infty$. Actually, one could add any interaction to the region $0 < r < \infty$ without changing the general form of the solution at infinity. For instance, suppose we add an interaction between the scalar test field and the gauge field of the vortex. 
Depending on the vortex field configuration, local coupling with the gauge field may affect the asymptotically defined constants $C_m$ and $d_m$. The phase-shift formula does not change by the gauge interaction, but its values are certainly modified by the factor $d_m$, which, like $C_m$, store information on the local interaction between the origin and infinity. Different models, like semilocal or Alice strings, shall present distinct non-minimal local interactions. The amount of this effect is hard to estimate without doing an explicit example, which, if one wants to tackle realistic gravitating vortices, requires numerical simulations. However, in the next section we give a simplified example mimiching this class of gravitating cosmic string background.
\\

Usually, in the partial wave approach, we rewrite the cosine in eq. (\ref{scalar_asymp_solution}) with complex exponentials, which gives
\begin{equation}
    \Phi_{solution} = \frac{1}{\sqrt{2\pi\lambda^\prime r}} \left[e^{-i \lambda^\prime r}\left(\sum_m C_m i^m e^{im\varphi} e^{-i\beta_{m^\prime}} \right) + e^{i\lambda^\prime r} \left(\sum_m C_m i^m e^{im\varphi}e^{i\beta_{m^\prime}} \right) \right].
    \label{solution_expanded}
\end{equation}
For the asymptotic ansatz we take
\begin{equation}
    \Phi_{ansatz} = f(\varphi) \frac{e^{i \lambda^{\prime}r}}{\sqrt{r}} + \sum_{m=-\infty}^\infty A_m i^m J_{m^\prime} (\lambda^\prime w)e^{im\varphi} = f(\varphi) \frac{e^{i \lambda^{\prime}r}}{\sqrt{r}} + (e^{i\lambda^\prime r \cos\varphi})_{mod}
    \label{scalar_ansatz}
\end{equation}
One of the reasons for writing the above equation in this form is that when $A_m = 1$, $a = 1$, $b = 1$ and $c = 0$ we recover the usual asymptotic ansatz known in quantum mechanics. In contrast to the standard approach, here $A_m$ is left to be determined based on the form of the solution at infinity, i.e. after the scattering.
\\

Now, let us express the asymptotic form of the ansatz (\ref{scalar_ansatz}) with plane waves, precisely as we did for the actual solution (\ref{solution_expanded})
\begin{align}
    \Phi_{ansatz} &= \frac{1}{\sqrt{2 \pi \lambda^\prime r }} \left[ 
    e^{-i\lambda^\prime r} \left(\sum_{m=-\infty}^\infty  A_m i^m e^{-i \frac{\lambda^\prime c}{b}} e^{i\alpha_{m^\prime}}e^{im\varphi}  \right) +\right.\nonumber\\ 
   & \left. +e^{i\lambda^\prime r} \left( \sqrt{2 \pi \lambda^\prime} f(\varphi) + \sum_{m=-\infty}^\infty  A_m i^m  e^{i \frac{\lambda^\prime c}{b}} e^{-i\alpha_{m^\prime}} e^{im\varphi}\right)   \right].
    \label{ansatz_expanded}
\end{align}
Comparing the coefficients of $e^{-i \lambda^\prime r}$, we obtain

\begin{equation}
  A_m e^{-i (\beta_{m^\prime} - d_m)} = C_m e^{-i \beta_{m^\prime}} \rightarrow A_m = C_m e^{-i d_m(\lambda)}.
\label{coefficient_ansatz}
\end{equation}
Comparing the coefficients of $e^{i \lambda^\prime r}$ and considering eq. (\ref{coefficient_ansatz}), results in

\begin{equation}
    f(\varphi) = \frac{1}{\sqrt{2\pi i \lambda^\prime}}
    \sum_{m=-\infty}^\infty D_m \left[e^{2id_m(\lambda)} - 1 \right]e^{im(\varphi - \delta \varphi)} = \sum_{m=-\infty}^\infty f_m(\varphi),
    \label{scalar_SA}
\end{equation}
where 
\begin{equation}
D_m \equiv C_m e^{i\left(\frac{\lambda^\prime c}{b} - d_m \right)},
\end{equation}
and $\delta \varphi = \frac{\pi}{2} \left(\frac{1}{b} - 1 \right) \ge 0$ knowing $b \leq 1$. The solution (\ref{scalar_SA}) has the extra factors $D_m$ and $\delta \varphi$ when compared to the result in quantum mechanics. However, in the limit $a \rightarrow 1$, $b \rightarrow 1$ and $c \rightarrow 0$ or equivalently, $\lambda' \rightarrow \lambda$, $m' \rightarrow m$ and $w \rightarrow r$ where one recovers the flat spacetime metric with the same parametrization in both limits, $r \rightarrow 0$ and $r \rightarrow \infty$, the extra factors dissappear and the results match.\

Now suppose $A_m = 1$, as it is in the standard quantum mechanics partial-wave approach. In this situation, the scattering amplitude is given by

\begin{equation}
f(\varphi) = \frac{e^{i\lambda' c/b}}{\sqrt{2 \pi i \lambda'}} \left ( \sum_{m} e^{2 i d_m}e^{i m (\varphi - \delta\varphi)}  - \delta(\varphi - \delta\varphi) \right ).
\end{equation}
It means that the conical structure produces a delta contribution to the scattering amplitude in the standard QM approach. This divergence was also found in \cite{deser1988classical}. In our approach, with $A_m$ free to be determined by the field solution at infinity, the scattering amplitude becomes

\begin{equation}
f(\varphi) = \frac{e^{i\lambda' c/b}}{\sqrt{2 \pi i \lambda'}} \left ( \sum_{m} C_m e^{idm} - \sum_{m} C_m e^{-i d_m} e^{im(\varphi - \delta\varphi)} \right ),
\end{equation}
which does not have any delta contribution due to the nontrivial mode dependent constants $C_m$ and $d_m$. Therefore, our formalism avoids the delta-function divergence.\par
Let us pause and connect our approach with the one in \cite{deser1988classical} for the particle scattering in a conical geometry. In their scenario, the whole space is conical and analytically determined. Because of a non-vanishing deficit angle, the scattering amplitude $f(\varphi)$ is singular, taking the standard partial wave expansion ansatz in quantum mechanics. They circumvented this problem by modifying the second term in (\ref{scalar_ansatz}) to match the solution determined at $r \rightarrow \infty$. Our reasoning is similar. Due to the conical structure at $r \rightarrow \infty$, we need to leave an extra degree of freedom to be fixed with the asymptotic solution after the scattering.
In contrast with the result in \cite{deser1988classical}, one needs to find the parameters in the extra factor $D_m$ numerically since our formalism deals with a class of spacetimes typically too complex to have a metric in a closed-form due to the matter-gravity interaction. Similar to \cite{deser1988classical}, here, the optical theorem is not satisfied, although there is no problem with the unitarity. The probability is conserved since the probability current is divergenceless. From this, we see that in this class of spacetimes, the optical theorem is no longer suitable to determine particle conservation.

Also, one can see that, due to conical structure, the angular variable in (\ref{scalar_SA}) has a deficit. In fact it is interesting to notice that $\delta\varphi$ is related to the angular difference between geodesics in the ideal string background \cite{anderson2015mathematical}, $\Delta\varphi = \frac{8\pi \mu}{b}$, $b=1-4 \mu$,  by the following relation
\begin{equation}
    \delta\varphi = \frac{1}{4} \Delta\varphi.
\end{equation}
The appearance of the deficit in the angular part of the scattering amplitude originates from the spacetime's conical structure with a deficit angle equal to $\delta=2\pi (1-L^\prime(\infty))$ \cite{Antonio_Eugenio}.\

Now, let us take a closer look at the phase shift. As we have shown before, the phase shift of the scalar field scattering in a gravitating cosmic string spacetime is given by
\begin{equation}
    \delta_m(\lambda) = \beta_{m^\prime} + \alpha_m = 
    \lambda^\prime \frac{c}{b} + \frac{m\pi}{2}\left(1 - \frac{1}{b} \right) + d_m(\lambda),
    \label{scalar_phaseshift}
\end{equation}
in which the second term matches the result found in \cite{deser1988classical} but with the addition of the first term accounting for the rescaling of the radial coordinate, and the last one accounting for the nontrivial spacetime configuration. As we mentioned before, the last term depends on the gravitating cosmic string model under study. An exciting feature of this formula is that the first term is already in the form of an approximation for the low-energy scattering. In this situation, the isotropic mode, $m=0$, is the only one to significantly contribute to the scattering amplitude since for other modes, a portion of the incident flux is going to be in the $\hat{\varphi}$-direction \cite{LeBellac}. In quantum mechanics, we define the scattering length, $l_{sc}$, by 
\begin{equation}
l_{sc} = \lim_{\lambda^\prime \rightarrow 0} \left | \frac{\delta_0(\lambda)}{\lambda^\prime} \right|.
\label{scat-length}
\end{equation}
For instance, when considering the potential of a hard sphere of radius $R_0$, the scattering length is equal to $l_{sc} = R_0$ as it should. Following (\ref{scat-length}), the scattering length of the gravitating cosmic string can be estimated as

\begin{equation}
    l_{sc} = \frac{c}{b} + \frac{d(d_0)}{d\lambda^\prime}(\lambda^\prime = 0),
\end{equation}
since for $\lambda^\prime = 0$ there is no scattering, resulting in $d_0(\lambda^\prime = 0) = 0$.

Now, knowing how to find the scattering amplitude, $f(\varphi)$, we can return to the computation of the differential cross-section. As the incoming and outgoing momenta, $\lambda$ and $\lambda'$, are not in general the same, the differential cross-section is given by \cite{Zettili}

\begin{equation}
    \frac{d\sigma}{d\varphi} = \frac{\lambda^\prime}{\lambda} |f(\varphi)|^2
    \label{scalar_difcross},
\end{equation}
ignoring the z-axis due to the symmetry of the system in this direction.  Clearly, if one includes the $z$-direction, the total cross-section diverges. 
\\
By sustituting eq. (\ref{scalar_SA}) into eq. (\ref{scalar_difcross}) and then integrating (\ref{scalar_difcross}) in $\varphi$, we obtain

\begin{equation}
    \sigma = \frac{4}{\lambda} \sum_{m=-\infty}^\infty |C_m|^2 \sin^2(d_m).
\end{equation}
This is exactly the known result in quantum mechanics with the extra term $|C_m|^2$. This term tends to 1 in the limit where one recovers the flat spacetime with the same parametrization when $r \rightarrow 0$ and $r \rightarrow \infty$.

\section{Toy model}
\label{sec6}
To show the utility of the new approach, we construct a simplified analytical model similar to a gravitating cosmic string background where it is possible to calculate the factors $C_m$ and $d_m$ explicitly. The metric is given by
\begin{gather}
\begin{aligned}
r < r_0 &: \quad N(r) = 1, \quad L(r) = r \\
r > r_0 &: \quad N(r) = a, \quad L(r) = br + c ,
\end{aligned}
\label{toymod_metric}
\end{gather}
which represents an empty cylinder of radius $r_0$ with all its energy concentrated on the zero-thickness walls. The spacetime is flat both inside and outside the cylinder, but the geometry outside is conical while the inside is Minkowskian. Imposing continuity of $L(r)$ at $r = r_0$ yields
\begin{equation}
r_0 = \frac{c}{1 - b},
\label{toymod_size}
\end{equation}
which suggests that the asymptotic conical geometry encodes information about the cylinder size. Here we use the approximation $r_0 \gg c/b = \mathcal{O}(1)$, which means $b$ is close to 1. Also, for simplicity, we set $a = 1$ to avoid any delta function in the field equation. Later, when tackling a more realistic model, we relax this condition. \par

The solutions to the field equations are
\begin{gather}
\begin{aligned}
r < r_0 &: \phi = e^{-iEt} e^{ikz} \sum_{m} i^m J_m(\lambda r) e^{im\varphi} \\
r > r_0 &: \phi = e^{-iEt} e^{ikz} \sum_{m} i^m C_m J_{m'}(\lambda w) e^{im\varphi}.
\end{aligned}
\end{gather}
If we take the asymptotic form of the solution outside ($r_0 \gg c/b$) when imposing the continuity of $\phi$ at $r = r_0$, we get

\begin{equation}
C_m \sqrt{\frac{2}{\pi \lambda r_0}} \cos(\lambda'r_0 + \beta_{m'}) = J_m(\lambda r_0).
\label{toymod_cond1}
\end{equation}
Now imposing the continuity of the derivative $d\phi/dr$ on the border of the tube results in
\begin{equation}
\frac{\lambda}{2} \Delta J_m(\lambda r_0) = C_m \sqrt{\frac{2}{\pi \lambda r_o}} \cos(\lambda r_0 + \beta_{m'}) \left[ \frac{1}{2r_0} + \lambda \tan(\lambda r_0 + \beta_{m'} )\right],
\label{toymod_cond2}
\end{equation}
where $\Delta J_m(x) = J_{m+1}(x) - J_{m-1}(x)$. Substituting \eqref{toymod_cond1} in \eqref{toymod_cond2}, we obtain

\begin{equation}
\tan(\lambda r_0 + \beta_{m'}) = \frac{1}{2} \left( \frac{\Delta J_m(\lambda r_0)}{J_m(\lambda r_0)} - \frac{1}{\lambda r_0} \right),
\label{toymod_cond3}
\end{equation}
which is equivalent to

\begin{equation}
d_m(\lambda) = \alpha_{m'} - \lambda\left(r_0 + \frac{c}{b} \right) + \tan^{-1}\left[ \frac{1}{2} \left( \frac{\lambda}{\lambda} \frac{\Delta J_m(\lambda r_0)}{J_m(\lambda r_0)} - \frac{1}{\lambda r_0} \right) \right].
\end{equation}
Inserting \eqref{toymod_cond3} in \eqref{toymod_cond1} gives the amplitude of the scattered field

\begin{equation}
C_m = \sqrt{\frac{\pi \lambda r_0}{2}} J_m(\lambda r_0) \left \{ 1 + \frac{1}{4} \left[ \frac{\Delta J_m(\lambda r_0)}{J_m(\lambda r_0)} - \frac{1}{\lambda r_0} \right]^2 \right \}^{1/2}.
\end{equation}

\begin{figure}[H]
\includegraphics[width=0.9\textwidth]{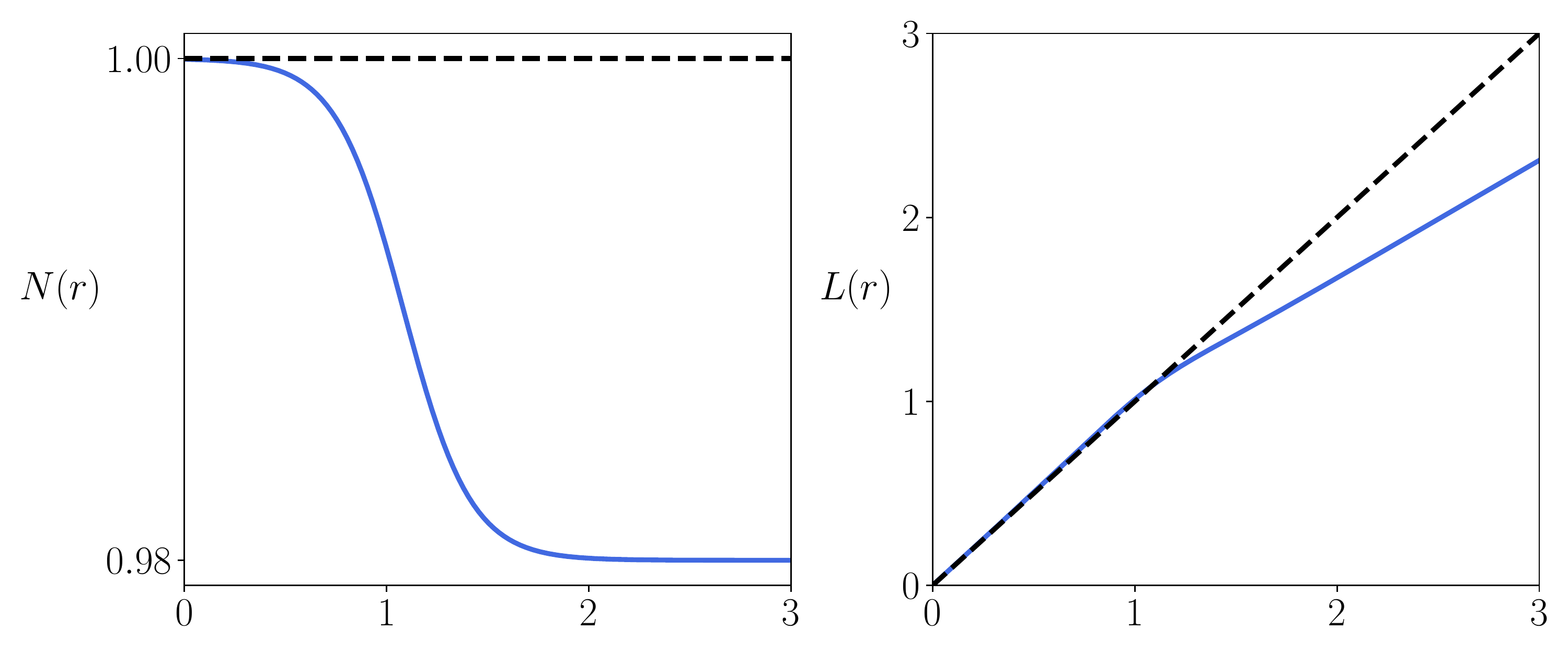}
\caption{Metric functions \eqref{toymod_metric_smooth} using $a = 0.98, b = 0.64, c = 0.39$ and $p = 3$. Dashed lines show the Minkowski counterparts.}
\label{fig:toymod_metric}
\end{figure}

It is worth mentioning that this toy model has its limitations. The metric \eqref{toymod_metric} has a curvature proportional to a delta function at $r = r_0$, so this is still a singular spacetime. Computing the cross-section we observe that its convergence is extremely slow. Therefore, in what follows, we develop a smooth version of the metric \eqref{toymod_metric}. Notice that the metric functions \eqref{toymod_metric} can be expressed using the Heaviside step function, $\Theta(x)$, in the following form
\begin{gather}
\begin{aligned}
N(r) &= \Theta(r_0 - r) + a \Theta(r - r_0), \\
L(r) &= r \Theta(r_0 - r) + (br + c) \Theta(r - r_0),
\end{aligned}
\label{toymod_metric_heaviside}
\end{gather}
which presents no transition region between Minkowski and the conical spacetime. We can create a smooth transition using an analytical approximation of the Heaviside function $\Theta(x)$, as follows
\begin{equation}
H(x) = \frac{1}{2} \left(1 + \tanh(px) \right), \quad H(x) \xrightarrow{p \to \infty} \Theta(x),
\label{toymod_Hfunc}
\end{equation}
which gives
\begin{gather}
\begin{aligned}
N(r) &= \frac{1}{2}\left \{ (a + 1) + (a - 1)\tanh\left[p(r - r_0)\right] \right \} \\
L(r) &= \frac{1}{2} \left \{ \left( (b + 1)r + c \right) + \left( (b - 1)r + c \right) \tanh \left[p (r - r_0) \right] \right \},
\end{aligned}
\label{toymod_metric_smooth}
\end{gather}
when combined with \eqref{toymod_metric_heaviside}.
One can see that $r_0$ can still characterize the vortex size in this toy model since the transition between the metric components occurs \emph{around} $r_0$. In fact, since this model is designed to mimic a class of gravitating cosmic string solutions, one can take \eqref{toymod_metric_smooth} as an approximation to realistic\footnote{Here, we use the term "realistic" for the metrics which are solutions of the Einstein field equations.} scenarios. In Figure \ref{fig:toymod_metric}, we plot the metric functions $N(r)$ and $L(r)$ compared with the Minkowski case.  \par

\begin{figure}[H]
\includegraphics[width=1.0\textwidth]{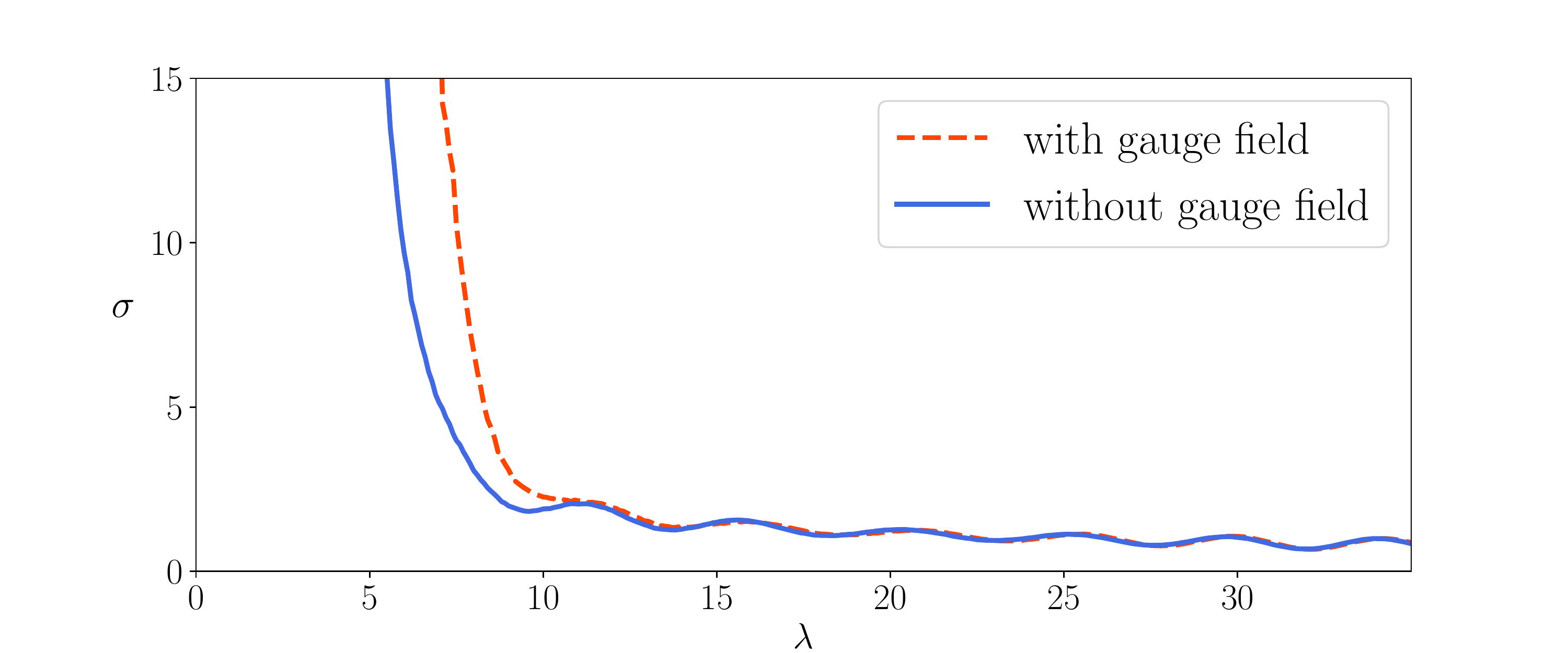}
\caption{The solid (dashed) line shows the total scattering cross-section of scalar field with $M = 1.0$ in the absence (presence) of the gauge field.}
\label{fig:toymod_sigma}
\end{figure}

One can substitute \eqref{toymod_metric_smooth} in \eqref{scalar_eom2} and solve the equation of motion numerically to extract the mode-dependent constants $C_m$ and $d_m$. The solid curve in Fig. \ref{fig:toymod_sigma} shows the scattering cross-section of a scalar field interacting with the background spacetime \eqref{toymod_metric_heaviside}.  Now, let us add an extra interaction in the aforementioned scenario. Consider the scalar test field interacting with the gauge field generating the vortex. We take the Nielsen-Olesen vortex \cite{nielsen1973vortex} with the gauge field given by
\begin{equation}
A^\varphi = \frac{n}{e r} \alpha(r) \hat{\varphi},
\end{equation}
where $e$ is the coupling constant between the gauge field and the scalar field of the vortex, $n$ is the winding number and $\alpha(r)$ satisfy $\alpha(r \to 0) \to 0$ and $\alpha(r \to \infty) \to 1$. The Nielsen-Olesen solution only depends on $n$ and $\beta = (m_s/m_v)^2$, where $m_s$ ($m_v$) is the mass of the scalar (gauge) boson \cite{vilenkin2000cosmic}. In \cite{Antonio_Eugenio}, the authors solved the Maxwell-Einstein equations for this vortex taking $n = 1$ and $\beta = 0.5$ and found conical parameters approximately around $a = 0.98, b = 0.64, c = 0.39$. We take the same conical and gauge field parameters, together with $p = 3.0$. In Figure \ref{fig:toymod_sigma}, one can observe how the gauge field coupling affects the total scattering cross-section of the scalar test field.

It is easy to notice that for large momentum (small wavelength), the scattering cross-section of the scalar field is much less pronounced compared with the low momentum (large wavelength) case for both scenarios, with and without the gauge field interaction, as expected. Only low momentum particles are affected significantly by the non-trivial background metric. However, in the presence of the gauge field interaction, the field is also scattered by the magnetic field of the vortex, leading to a larger scattering cross-section. For large enough momenta, the interaction with the local potentials becomes less relevant, resulting in a small scattering cross-section. As the momentum tends to infinity, the cross-section should vanish.

\section{Fermion field scattering}
\label{sec4}
In this section, we will show that with a simple adjustment, one can find the scattering differential cross-section for a Dirac particle interacting with a gravitating cosmic string background.
In this case, we treat each component of the spinorial field as a scalar field and find its associated differential cross-section. The total differential cross-section for the scattering of a fermionic particle is the average value of all components \cite{bazeia2018dirac}.\

The Dirac equation in a curved spacetime is in the form
\
\begin{equation}
(\gamma^A e^\mu \, _A \partial_ \mu - \gamma^A \Gamma_A + iM_f)\Psi = 0,
\label{Dirac-eq}
\end{equation}
where $\Gamma_A$ and $M_f$ are the spin connection and the fermion mass, respectively. The Greek indices are lowered and raised by the spacetime metric $g_{\mu \nu}=diag(N^2,-1,-L^2,-N^2)$ and the latin indices by $\eta_{AB}=diag(1,-1,-1,-1)$. When $r\rightarrow 0$ the positive- and negative-energy fermionic mode functions (\ref{Dirac-eq}) are given by \cite{de2013fermionic}
\begin{equation}
\psi^\pm_j(t, r \rightarrow 0, \varphi, z) \propto e^{\mp iEt} e^{ikz} e^{ij\varphi}
\begin{pmatrix}
J_{\beta_j}(\lambda r)e^{-i\varphi/2} \\
s J_{\beta_j + \epsilon_j}(\lambda r)e^{i\varphi/2} \\
\pm \frac{k -i s \epsilon_j\lambda}{E \pm M} J_{\beta_j}(\lambda r)e^{-i\varphi/2} \\
\mp \frac{k -i s \epsilon_j\lambda}{E \pm M} J_{\beta_j + \epsilon_j}(\lambda r)e^{i\varphi/2}\\
\end{pmatrix},
\end{equation}
where $j = \pm 1/2, \pm 3/2, ...$, $\epsilon_j = sgn(j)$, $s=\pm 1$ and $\beta_j = |j| - \epsilon_j/2$. The proportionality constant is to be defined based on the chosen initial condition, just as we did with the scalar field. Now, the solution, when $r \rightarrow \infty$, has a similar form

\begin{equation}
\psi^\pm_j(t, r \rightarrow \infty, \varphi, z) = C_j e^{\mp iEt} e^{ikz} e^{ij\varphi}\sqrt{\frac{2}{\pi \lambda' r}}
\begin{pmatrix}
\cos(\lambda^\prime w -\alpha_{\beta_{j^\prime}}+ d_{j}^{0}(\lambda)) e^{-i\varphi/2} \\

s \cos(\lambda^\prime w - \alpha_{\beta_{j^\prime} + \epsilon_{j^\prime}}+ d_{j}^{1}(\lambda)) e^{i\varphi/2} \\

\pm \frac{k - i s \epsilon_{j^\prime }\lambda'}{E \pm M^\prime} \cos(\lambda^\prime w - \alpha_{\beta_{j^\prime}} + d_{j}^{2}(\lambda)) e^{-i \varphi/2} \\

\mp \frac{k - i s \epsilon_{j^\prime}\lambda'}{E \pm M^\prime} \cos(\lambda^\prime w - \alpha_{\beta_{j^\prime} + \epsilon_{j^\prime}} + d_{j}^{3}(\lambda)) e^{i \varphi/2}\\

\end{pmatrix},
\end{equation}
where $j^\prime = j/b$, $w = r + c/b$, $M^\prime = aM$ and ${\lambda^\prime}^2 = (E/a)^2 - (k/a)^2 - M^2$. Notice that again, we have to determine the constants $C_j(\lambda)$ and $d^i_j(\lambda)$ based on the specific gravitating cosmic string model, and hence we are going to have four different phase shifts, depending on how each component of the fermionic field is affected by the non-trivial spacetime configuration.

Now, in order to apply our formalism to the fermionic field, one needs to treat each component as a scalar field and find its associated scattering amplitude $f^i(\varphi)$. The differential cross-section is then given by the average of the differential cross-sections of all components as follows
\begin{equation}
\frac{d\sigma}{d\varphi} = \frac{1}{4} \frac{\lambda^\prime}{\lambda}\sum_{i = 0}^{3} | f^i(\varphi) |^2.
\end{equation}

\section{Conclusion}
\label{sec5}
In this work, we have developed a new approach to relativistic quantum scattering in a class of models regarding gravitating cosmic string spacetimes. Our approach is a modification of the usual partial wave expansion, which was motivated by the formalism in \cite{deser1988classical} for a particle scattering in a conical geometry.\
In the first part, we considered the Klein-Gordon equation non-minimally coupled with gravity. The background metric comes from a gravitating cosmic string spacetime in the form of an extended vortex. This class of spacetimes is cylindrically symmetric with the general line element given in eq. (\ref{metric_general}) where the metric components have the forms (\ref{metric_zero}) and  (\ref{metric_infty}) when $r\rightarrow 0$ and $r \rightarrow \infty$, respectively. These forms reflect the fact that the background spacetime close to and far from the vortex center is flat and generally with different parametrization. However, in between, $0<r<\infty$, the curvature of the spacetime is non-zero in general and can have a complicated form. The solution to the Klein-Gordon equation is a linear combination of the Bessel functions, parametrized with mode number $m$, at the center of the vortex. Far from the vortex, the solution is a linear combination of Bessel and Neumann functions but with the momentum eigenvalue corrected to account for the conical structure. We have found the resulting phase shift of the scalar field interacting with the background spacetime by taking the asymptotic forms of the above special functions for $r\rightarrow 0$ and $r \rightarrow \infty$. We have shown that, besides the contribution from the conical structure of the spacetime, which is the same as the case of a pure conical geometry studied in \cite{deser1988classical},  the phase shift includes a contribution originating from the complex spacetime geometry in $0<r<\infty$. The different spacetime parametrization resulting in different momenta, $\lambda$ and $\lambda'$, also affects the phase shift. 

It has already been shown that in conical spaces, the scattering problem is nontrivial. As shown initially in \cite{deser1988classical}, the partial wave approach needs to be modified in these scenarios. Here we have shown how to modify it in a gravitating cosmic string spacetime. We left a free parameter in the unscattered wave and then fixed it by the field solution at $r \rightarrow \infty$. This procedure is shown to remove singularities off the scattering amplitude and yields a modified form of the partial wave expansion, with an extra factor compared with the canonical form in quantum mechanics. However, as expected, one recovers the canonical form when the flat spacetime has the same parametrization for $r\rightarrow 0$ and $r \rightarrow \infty$. In addition, we presented a toy model in order to calculate the factors $C_m$ and $d_m$ analytically.\par 

We have also discussed how to implement other interactions besides gravity into the formalism. The effect of any new local interaction enters our approach via its effect on the amplitude, $C_m$, and phase, $d_m$, of the field at infinity. For instance, the interaction with the gauge field that creates the vortex will undoubtedly affect the cross-section, but the amount of change is model-dependent. We have added the interaction of the field with the Nielsen-Olesen vortex gauge field in the toy model and discussed the interplay between the spacetime curvature and the gauge field configuration in the scattering cross-section.

Finally, we have shown that the new partial wave approach is also valid for the Dirac field in the same spacetime background, applying our scalar field scattering formalism and the consequent scattering differential cross-section for each component. This way, taking the average value gives the total differential cross-section for a fermionic field.

\section*{Acknowledgments}

AM acknowledges the financial support from the Brazilian agencies CAPES and CNPq, Grant Number 309368/2020-0, besides the financial support from Universidade Federal de Pernambuco Edital Qualis A. MS also thanks the Brazilian agency CNPq for the financial support.

\printbibliography

\end{document}